\documentclass[usenatbib]{mn2e}
\usepackage{graphicx,amsmath,multirow,amssymb}
\usepackage{subfig}
\usepackage{natbib}
\usepackage[usenames,dvipsnames]{color}

\newcommand{\comment}[1]{}

\def\simgt{\lower.5ex\hbox{$\; \buildrel > \over \sim \;$}}
\def\simlt{\lower.5ex\hbox{$\; \buildrel < \over \sim \;$}}

\newcommand\aj{{AJ}}%
\newcommand\apj{{ApJ}}%
\newcommand\apjl{{ApJ}}%
\newcommand\aap{{A\&A}}%
\newcommand\mnras{{MNRAS}}%
\newcommand\nat{{Nature}}%

\title[The extreme He-rich population of NGC 2808]{Self-enrichment in Globular Clusters:\\
the extreme He-rich population of NGC 2808}
\author[Di Criscienzo et al.]{Di Criscienzo, M.$^1$, Ventura$^1$ P., D'Antona$^1$ F., Dell'Agli$^{2,3}$, F. \&  Tailo, M. $^{4}$\\
$^{1}$INAF -- Osservatorio Astronomico di Roma, Via Frascati 33, 00078, Monte Porzio Catone (RM), Italy \\
$^{2}$Instituto de Astrof\'{\i}sica de Canarias, E-38205 La Laguna, Tenerife, Spain \\
$^{3}$Departamento de Astrof\'{\i}sica, Universidad de La Laguna (ULL), E-38206 La Laguna, Tenerife, Spain\\
$^{4}$Dipartimento di Fisica e Astronomia `Galileo Galilei', Univ. di Padova, Vicolo dell'Osservatorio 3, I-35122 Padova, Italy
}

\begin{document}

\date{Accepted, Received; in original form }

\pagerange{\pageref{firstpage}--\pageref{lastpage}} \pubyear{2018}

\maketitle

\label{firstpage}

\begin{abstract}
Almost several decades after the discovery of the first multiple populations in galactic 
globular clusters (GC)  the debate on their formation is still  extremely current and 
NGC2808 remains one of the best  benchmark to test any scenario for their origin and the 
evolution. In this work we focus on the chemical composition 
of stars belonging to the extreme He-rich population populated by stars with the most 
extreme abundance of Mg, Al, Na, O and Si. We checked whether the most recent measures are 
consistent with the AGB yields of stars of $6.5-8~M_{\odot}$. These stars evolve on time 
scales of the order of 40-60 Myr and eject matter strongly enriched in helium, owing to a 
deep penetration of  the surface convective zone down to regions touched by CNO 
nucleosynthesis occurring after the core He-burning 
phase. Since the big unknown of the AGB phase of massive stars is the mass loss, we propose a new approch 
that takes into account the effects of the radiation pressure on dust 
particles. We show that this more realistic description is able to reproduce the observed 
abundances of Mg, Al, Na  and Si  in these extreme stars. The large spread in the oxygen 
abundances is explained by invoking deep mixing during the RGB phase. 
It will be possible to check this work hypothesis  as soon as the oxygen measurements 
of the main sequence stars of NGC2808 will be available.

\end{abstract}


\section{Introduction}
The recent decades have witnessed a growing interest towards the evolution of Globular
Clusters (GC). Results from high resolution spectroscopy, photometry and 
spectrophotometry, have shown that Galactic GC harbour a variety of stellar populations, 
differing in their chemical composition. This is at odds with the classic, traditional 
assumption, that GC are simple stellar populations, composed by coeval and chemically 
homogeneous stars.

All the GC of the Milky Way so far examined present star-to-star variations, which trace 
well defined abundance patterns, such as the C-N and O-Na anti-correlations, whose extension 
varies from cluster to cluster \citep{gratton12}. These abundance variations were observed 
also at the surface of unevolved stars \citep[e.g.][]{gratton01}; for these stars, unlike 
red giants, can be disregarded the effects of any possible `in situ' production mechanism able to reproduce at the same time  all the chemical patterns.
A plausible interpretation of  these 
observations is that in GC a variety of generations
of stars coexist, each characterized by a different chemical composition. 
This conclusion is further reinforced by the observations indicating that some GC also exhibit 
a Mg-Al trend \citep{kraft97,gratton01,carretta09,meszaros15}, involving two species, Mg and Al, whose surface abundance  remains unchanged during the red giant branch (hereinafter RGB) evolution, even in case of 
deep mixing.

NGC 2808 has so far played a pivotal role in the ongoing debate regarding the modality
with which new generations of stars formed in GC, after the birth of the original population.
The great interest towards this cluster is motivated by the complex morphology of the horizontal
branch (HB) and the peculiar distribution of stars across the main sequence (MS). The HB 
has a clumpy structure, with some gaps in the distribution of 
stars in the colour-magnitude diagram (CMD) \citep{bedin00}; the MS exhibits 
a significant spread in colour, and is split into three discrete sequences 
\citep{dantona05, piotto07}.

These results indicate the presence of three groups of stars, 
differing in the helium content, $Y$: a) the primordial population, with $Y=0.25$, which 
correspond to the stars on the red side of the HB and on the red main sequence; b) the 
bluest and faintest objects in the HB, the counterparts of the stars populating the 
blue MS, with $Y>0.35$; c) the "intermediate" population, with helium 
$0.27 < Y <0.30$ \citep{dantona04, dantona05, piotto07}. 
\citet{bragaglia10} performed a chemical tagging of two stars belonging, respectively, to 
the He-normal and He-rich main sequences, finding that the blue-MS star shows a 
huge enhancement of N, a depletion of C, an enhancement of Na and Al, and a small depletion 
of Mg, with respect to the red MS star. This is exactly what is expected if stars on the 
blue MS formed from the CNO-processed ejecta produced by an earlier stellar generation.

The recent analysis by \citet{milone15}, based on HST UV data from the Legacy Survey of
Galactic globular Clusters \citep{piotto15}, outlined an even more complex stellar 
population, suggesting that NGC 2808 harbours 5 groups of stars, differing in their 
chemical composition.

\citet{dantona16}, in the context of the scenario based on
self-enrichment by massive AGB stars \citep{ventura01, dercole08}, proposed a temporal 
sequence for the formation of the various populations in NGC 2808. 
In particular  the stars belongin to population E, according to the classification by Milone et . (2015), formed 40-60 Myr after the formation of the cluster, directly  from the ejecta of AGB stars of mass  7-8 Mo, after the end of SNII that means with no dilution with pristine gas.
Because the ejecta of massive  AGB stars are enriched in helium, we know from stellar evolution theories that
stars born with this chemical composition would populate the blue side of the HB during 
the core helium burning phase \citep{dantona02} and would define a blue MS during the core 
hydrogen burning evolution; this would explain the peculiar distribution of NGC 2808
stars in the CMD. 

Assessing the reliability of the framework proposed by the afore mentioned works can be
considered as a general test for the AGB self-enrichment scenario itself. NGC 2808 is by 
far the most complex GC so far investigated: the capability to reproduce all the observations 
collected so far would be a strong point in favour of any theoretical schematization.
The present investigation  falls within this context.
The recent high-resolution spectroscopy results by \citet{carretta18} allow a significant 
step forward towards this direction. This work, coupled with the results by 
\citet{carretta15}, presents data regarding all the light elements, including Mg, Al and 
Si, whose abundances were not considered in the analysis by \citet{dercole10} and 
\citet{dantona16}. 

In this paper we want to test the scenario proposed by \citet{dantona16} for the star formation
history of NGC 2808. We focus on the chemical composition of stars belonging to the 
population E, to check whether their observed surface abundances are consistent with the
AGB yields of stars of $6.5-8~M_{\odot}$ (SAGB).

The paper is structured as follows: in section 2 we present an overview of the 
self-enrichment scenario by AGB stars, whereas section 3 focuses on the importance of
NGC 2808 in the current debate on the formation of multiple population in GC; in 
section 4 we present the SAGB models, calculated on purpose for this work, with the same
chemical composition of the stars belonging to primordial population of NGC 2808;
section 5 presents a comparison between the yields of massive AGB stars with the chemical
composition of the NGC 2808 stars in the group E. The conclusions close the paper and  are given in
section 6.

\section{Self-enrichment by AGB stars}
According to the AGB scenario for self-enrichment in globular clusters, the formation of
new stellar generations began after the epoch of type II SNe explosions, from the ashes of 
stars of mass $4~M_{\odot} < M < 8~M_{\odot}$, belonging to the first generation (FG)  
\citep{ventura01, dercole08}. During the AGB phase these objects lose the external 
mantle, releasing gas processed by hot bottom burning (HBB), which consists into
the proton-capture nuclear activity experienced at the 
base of the external envelope \citep{renzini81, blocker91}. 

The chemical composition of the stars belonging to the newly formed second generation
(SG) of the cluster is determined by the complex interplay among the metallicity, 
the dilution of the AGB gas with pristine matter, the epochs when 
dilution began and when the formation of the SG was halted by type Ia SN explosions
\citep{dercole11, vesperini13, dantona16, dercole16}.

The interpretation of the observed abundances of SG stars requires the
use of dilution curves, calculated by assuming mixing of AGB ejecta with various percentages
of pristine matter, with the same chemical composition of FG stars \citep{ventura08,
ventura09, ventura16, dantona16, flavia18}. This makes the understanding of the
observed chemical patterns more complicated, as the results obtained are sensitive to the 
efficiency of different physical phenomena, primarily HBB, dilution and mass loss.

\subsection{The signature of SAGB stars in Globular clusters with extreme populations}
A robust evaluation of the reliability of any self-enrichment scenario
and a straightforward reconstruction of the star formation history of the cluster
is possible when part  of the SG formed directly from the gas expelled by the 
polluters. In this particular case the AGB scenario can be tested by comparing the surface 
abundances of FG stars with the chemical composition of the AGB ejecta.
The clusters harbouring this class of stars can be easily identified, because
the presence of fully contaminated SG stars is associated to an
evident blue tail characterizing the HB morphology \citep{dantona02, dantona04},
determined by the large helium enrichment.

This approach was followed by \citet{marcella11, marcella15} and \citet{ventura16} to 
study the formation of multiple populations in NGC 2419 and by \citet{tailo15, tailo16} to
investigate the various stellar groups in $\omega$ Centauri; 
these are the other two Galactic clusters, together with NGC2808, which harbour an extreme 
SG, likely formed without dilution with pristine gas.

In the AGB self-enrichment scenario the formation of stars strongly enriched in helium 
from gas undiluted with pristine matter can occur within 10-20 Myr since the end of the
epoch of type II SNe explosions, from the winds of stars of initial mass in the range
$7-8~M_{\odot}$\footnote{The exact values of the mass depend  on the adopted metallicity and on the assumptions
regarding the extent of the extra-mixing from the border of the convective core during
the core H-burning phase. When no overshooting is considered, the above range in mass
would shift upwards to $9-10~M_{\odot}$}. These stars evolve on time scales of the order 
of 40-60 Myr and eject matter strongly enriched in helium, owing to a deep penetration of 
the surface convective zone down to regions touched by CNO nucleosynthesis, during the second
dredge-up (hereinafter SDU) episode, occurring after the core He-burning phase. The helium 
enrichment is higher the more massive is the star; the largest helium mass fractions,
$ Y \sim 0.37$, are achieved by stars of initial mass $\sim 8~M_{\odot}$
\citep{ventura10}.

An important feature of the evolution of this class of stars is the off-center ignition of 
carbon burning in condition of partial degeneracy. The consequent development
of a convective flame, propagating inwards, favors the formation of a degenerate core,
composed of oxygen and neon \citep{garcia94, garcia97, ritossa96, siess06, siess07, siess10}.
The thermally pulsing phase of these stars is commonly referred to as Super Asymptotic
Giant Branch (SAGB) evolution.

The pollution from SAGB stars in low-metallicity environments, such as GC, is not
completely understood. There is a general consensus that the gas ejected by these objects 
is enriched in helium, because most of the helium increase at the surface takes place in 
the evolutionary phases previous to the beginning of thermal pulses (hereianfter TP). 

Regarding the other chemical species, particularly the light elements involved in the 
correlation/anti-correlation patterns observed in GC stars, the yields are sensitive
to the details of modelling of the TP phase, particularly to the 
description of mass loss, the efficiency of mixing at the base of the envelope and
the strength of HBB. An exhaustive discussion on this argument, with the role played by
different physical mechanisms, is given in \citet{doherty14} and it will be investigated in the next section.

\section{NGC 2808: a cornerstone in the understanding of multiple populations in GC}
\label{2808}


In the context of the self-enrichment process by AGB stars, \citet[hereafter D16]{dantona16} proposed
a temporal sequence for the formation of the five populations belonging to NGC 2808 (see Fig.1), which 
can be summarized as follows: a) the FG of the cluster defines the group P1(B in D16); b) regarding 
the SG, the stars with the most extreme chemical composition (group E) formed within 
$\sim 20$ Myr from the end of the epoch of type II SNe explosions, directly from the gas 
ejected by SAGB stars; c) the remaining SG stars (groups I2, I1 and P2) formed later, from 
the gas lost by AGB stars of lower mass ($4-6~M_{\odot}$), diluted with pristine matter.

For the reasons discussed in the previous section, the reliability of this understanding 
can be assessed via the analysis of the chemical composition of stars in group E:
because no dilution occurred when these stars formed, it is crucial that the SAGB yields
correspond to the chemical patterns traced by these objects. Compared to
D16 we can make a significant step forward in this direction, because
that work  does not include the discussion of the abundances of
aluminum and silicon. These elements are of paramount importance to interpret the
observations, because, unlike CNO elements and sodium, their surface abundance does
not change during the RGB evolution, thus the observations reflect the original chemical
composition of the stars.

\begin{figure}
\resizebox{1.0\hsize}{!}{\includegraphics{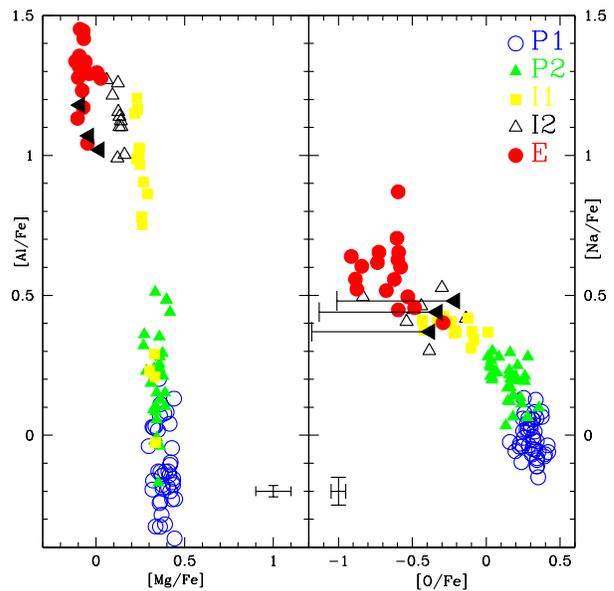}}
\caption{Abundance ratios [Al/Fe] and [Na/Fe] as a function of [Mg/Fe] and [O/Fe] 
respectively in RGB stars of NGC 2808 measured by Carretta et al. (2015) and Carretta 
et al. (2018). Different colours indicate the five populations as defined in Carretta 
(2015) using the Mg-Na plane. Large and black triangles are the yields computed in this work  
for AGB stars of M=6.5, 7.0 and 7.5 $M_{\odot}$ (see also Table 1). In the case of Oxygen 
the uncertainties related  the adopted extra-mixing (see text) is also displayed. }
\label{fanti}
\end{figure}

\citet{carretta18} show that group E stars exhibit the most
extreme chemical composition, for all the species observed. Indicating with 
$\delta X=[X/Fe]_{E}-[X/Fe]_{FG}$ the differences between the surface mass fraction of
the species $X$ for group E and FG stars (P1 according their nomenclature), 
\citet{carretta18} (see Table 6 and Fig. \ref{fanti}) find $\delta$ Na $\sim$ +0.6, 
$\delta$ Mg $\sim$ -0.4, $\delta$ Al $\sim$ +1.5, $\delta$ Si  $\sim$ +0.2.

The situation for oxygen is more complex, because the observations indicate for E stars 
$-1.5 < \delta O < -0.7$, an excursion which is significantly larger than the expected 
scatter ($\sim 0.1$) and also  larger than the one observed in the FG. We will discuss this point in the next section.

These recent results make appropriate a  revison of the scenario given in D16 and/or the SAGB models used in that work.
First of all   in the SAGB models by \citet{ventura13} used by D16 to interpret the abundance patterns of E stars in NGC 2808, with the sole exception of helium, the most contaminated gas is produced by stars of initial mass 
$\sim 6~M_{\odot}$ while the SAGB yields of O, Mg, Al and Si are not significantly  altered with respect to the original chemistry (see Fig.~6 in Ventura et al. 2013).
In particular D16 remarked that the \citet{ventura13} models correctly predict Mg depletion in group E, but their Mg abundances are $\sim$0.15\,dex larger than those observed, as recently noticed also by \cite{carretta18} so further exploration is needed to re-evaluate the Mg case.\\ 
They also suggested that the small oxygen measured in same E stars could be explained with deep mixing during the RGB phase, an hypothesis that will be tested in this work through the computation  of specific models.


In the following section we present new SAGB models, calculated on purpose to
explain stars belonging to group E in NGC 2808, we will check in this work  to check whether the models are
consistent with the observations.

\begin{figure*}
\resizebox{0.45\hsize}{!}{\includegraphics{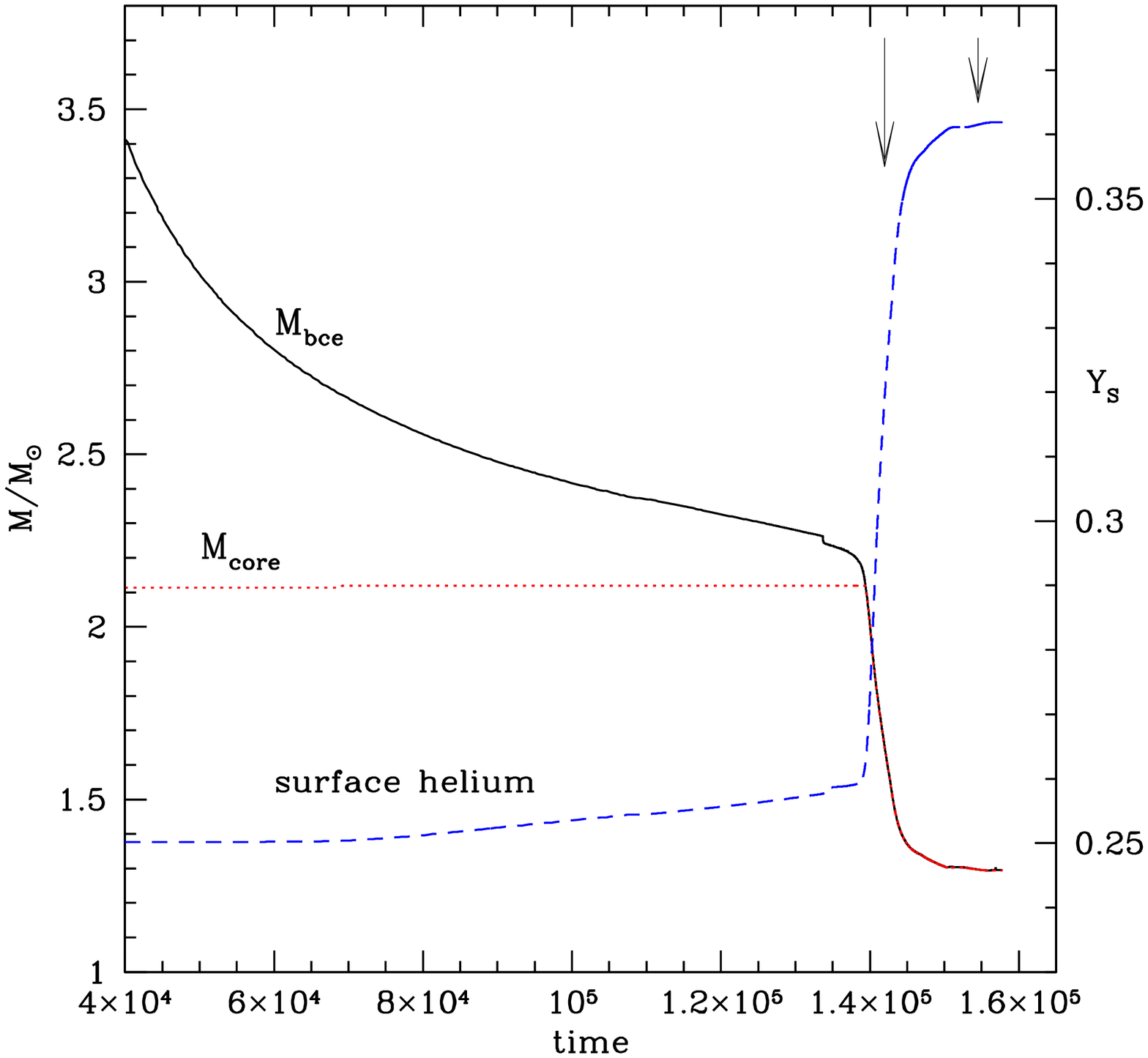}}
\resizebox{0.45\hsize}{!}{\includegraphics{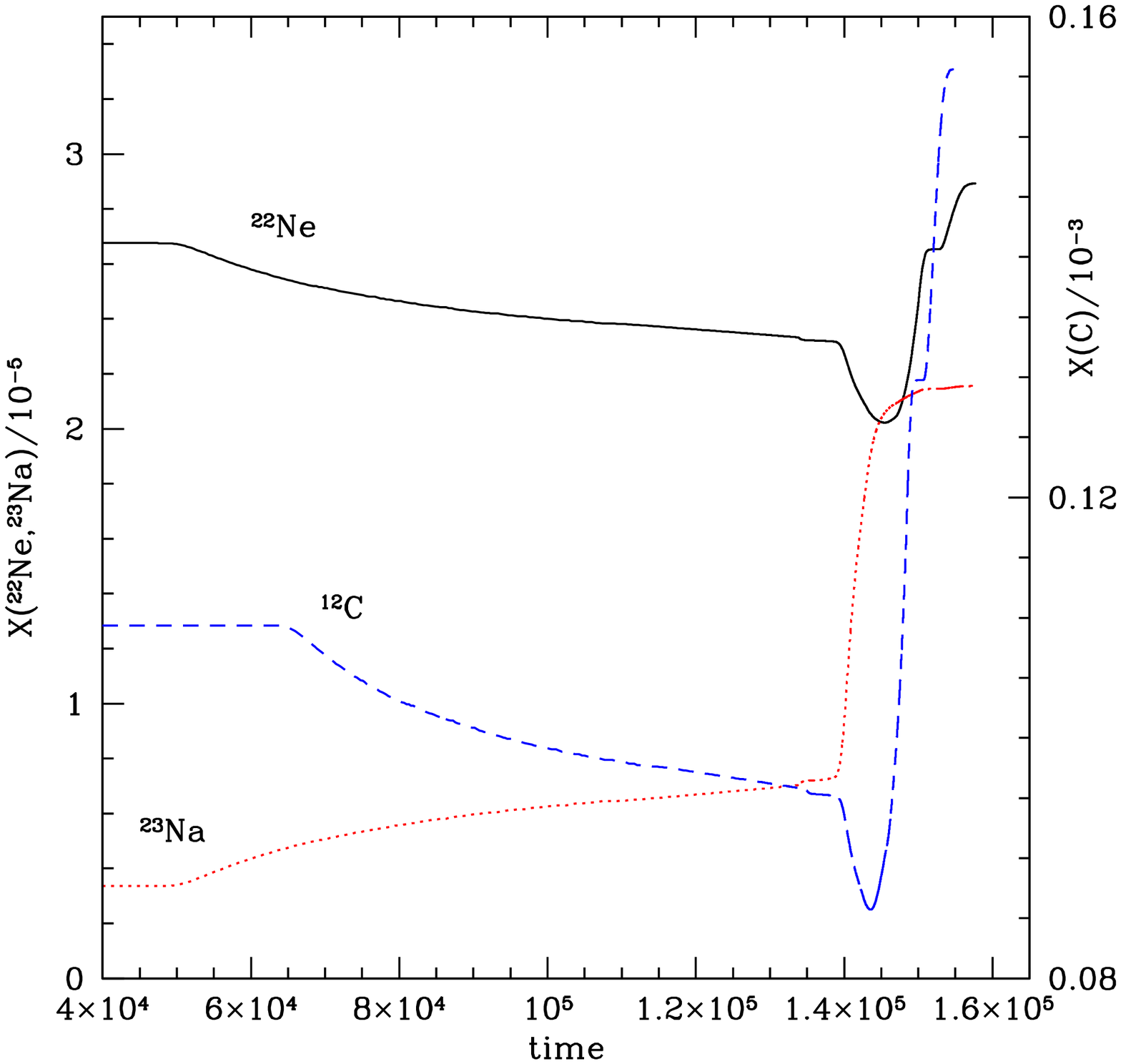}}
\caption{Left: The evolution of the mass of the base of the envelope (solid, black line)
and of the core mass (dotted, red) of a $7.5~M_{\odot}$ model, during the evolutionary 
phases following the exhaustion of central helium and before the beginning of thermal pulses. 
Times are counted since the end of central helium. The vertical arrows 
indicate the times of the ignition of the first episode of carbon burning (arrow on the left)
and of the formation of the O-Ne core (right). The dashed, blue track indicate the surface 
helium abundance (scale on the right). Right: The evolution of the surface mass fraction 
of $^{22}$Ne (solid, black lines) and sodium (dotted, red) in the same evolutionary phases 
shown in the left panel. The behaviour of the surface carbon (dashed, blue line, scale on 
the right) is also shown.}
\label{f2dup}
\end{figure*}

\section{SAGB modelling}
\label{input}
\subsection{Chemical and physical input}
To study the extreme population of E stars in NGC 2808 we calculated ad hoc SAGB models
with initial mass in the range $6.5-8~M_{\odot}$. The chemical composition is the 
same as the FG of the cluster, i.e. the stars belonging to group P1 in 
\citet{carretta18}. The initial abundances of O, Na, Mg, Al and Si,
reported in Table.~\ref{tabyield}, were chosen accordingly. For carbon and nitrogen we 
assumed solar-scaled abundances, with $[C/Fe]=[N/Fe]=0$. With these choices, the 
metallicity corresponding to the average $[Fe/H] \sim -1.15$ measured by \citet{carretta18} 
is $Z = 2\times 10^{-3}$. The initial helium was taken as $Y=0.25$.

We used the same physical ingredients described in detail in
\citet{ventura09}. We briefly recall here the most relevant input.\\

{\it Convection}. Convection was modelled according to the Full Spectrum of Turbulence 
(hereinafter FST)
model, developed by \citet{cm91}. Nuclear burning and mixing of chemicals are treated
simultaneously, according to the diffusive schematization by \citet{cloutman}. During the core
burning phases we assumed overshoot from the border of the convective core, modelled
via an exponential decay of the convective velocities within the radiatively stable 
zones, with an e-folding distance $\zeta=0.02 \times H_P$ \citep{ventura98}. 
No extra-mixing was assumed during the TP phase.\\

{\it Nuclear reactions}. We used the NACRE compilation \citep{angulo99} for the 
cross sections of the various reactions included in the nuclear network. The exceptions 
are: $^{14}$N$(p,\gamma)^{15}$O
\citep{formicola04}; $^{22}$Ne$(p,\gamma)^{23}$Na \citep{hale02}; 
$^{23}$Na$(p,\gamma)^{24}$Mg and $^{23}$Na$(p,\alpha)^{20}$Ne \citep{hale04}. For
what attains the rate of the proton capture reactions by the heavy Mg isotopes we used
the same input as in \citet{ventura18}, which allowed us to reproduce the Mg-Al trends 
observed in M13, NGC6752 and NGC2419.\\

{\it Mass loss}. 
We follow the description by \citet{blocker95} to model mass loss during the TP
phases. This treatment provides a numerical fit of $\dot M$ as a function of mass and
luminosity of the star, based on hydrodynamic models of O-rich envelopes, in which mass
loss is driven by radiation pressure on the dust particles formed in the circumstellar 
envelope. In agreement with out previous works, in the initial TP phases we
assume the free parameter entering the \citet{blocker95} formula $\eta_R=0.02$, in 
agreement with the calibration based on the luminosity function of Li-rich stars in the 
Magellanic Clouds, given in \citet{ventura00}. However, in the present investigation 
we propose an innovative approach, that considers the change in the amount of dust formed,
as a consequence of the alteration of the surface chemical composition of the star.

In the circumstellar envelope of O-rich, AGB (and SAGB) stars, the most relevant dust
species are alumina dust and silicates \citep{fg06, ventura14}. Under HBB conditions,
the presence of these particles, owing to the large luminosities experienced, 
triggers a significant acceleration of the stellar wind,
even in metal-poor environments \citep{marcella13}. 

Because alumina dust is extremely transparent to the electromagnetic radiation, the acceleration
of the wind is mainly due to silicates and, more specifically, to iron-free olivine, 
a compound including magnesium, silicon and oxygen atoms, whose formula is Mg$_2$SiO$_4$. 

The reaction leading to the formation of iron-free olivine is \citep{fg06}

$$
2Mg + SiO + 3H_2O \longrightarrow Mg_2 SiO_4 + 3H_2
$$

Given the extreme stability of the CO molecules, only the oxygen not locked into CO 
is available to form dust; therefore, the quantity of dust which forms depends also on 
the number of carbon particles. The rate with which the reaction above proceeds is 
determined by the addition of Mg atoms, or by the addition of SiO molecules, or by the 
supply of oxygen by water molecules. Therefore, the rate of the silicates growth is 
proportional to the least (in number) among  magnesium, silicon and oxygen, which is 
called "key element".

In the present work we kept $\eta_R$ constant until the number density of the key element 
is unchanged.
We will see in section 4.4 that this holds only in the initial phases, before the destruction of
magnesium determine a significant decrease in the rate of dust formation and for this reason we assumed
that $\eta_R$ scales with the abundance of the key element after the beginning of Mg destruction till the achievement of the C-star phase after which we kept the mass loss rate constant until the consumption of the whole envelope.
Note that the above correction is required only when extremely strong HBB conditions are activated; in stars of lower mass and, more generally, in higher metallicity stars, the surface magnesium is not consumed severely, thus dust is formed efficiently for the whole TP evolution.



\subsection{The pre-TP evolution}
\label{pretp}
The evolutionary phases following the end of core helium burning and before the
beginning of TP of stars undergoing the SAGB evolution are characterized by the
inwards penetration of the convective envelope and the ignition of carbon burning.

The left panel of Fig.~\ref{f2dup} shows the variation of the position of the base of the
surface convective zone and of the helium-hydrogen interface in the model of (initial)
mass $7.5~M_{\odot}$. In the final part of this phase the base of the envelope 
reaches the chemical discontinuity, starting the second dregde-up (SDU); the surface 
convection pushes the H/He boundary inwards, by $\sim 1~M_{\odot}$, and reaches stellar 
regions enriched in helium and, more generally, contaminated by nuclear activity. 

As indicated by the left arrow in the figure, the first ignition of carbon burning 
occurs shortly after the beginning of the SDU. The formation of the O-Ne core (right arrow),
towards the end of the second ignition of carbon, takes place when the SDU is practically
finished. 

We see that the occurrence of SDU determines a significant increase in the surface helium,
whose mass fraction is raised by $0.11$. This result is extremely robust, because even 
in case of a more 
efficient SDU the increase in the surface helium would remain substantially unchanged: 
indeed in this case the base of the envelope would reach layers of the star enriched in 
carbon, where helium burning occurred. The present results are in agreement with the 
investigation by \citet{doherty14}, who studied the SAGB evolution of stars of similar
metallicity.

The right panel of Fig.~\ref{f2dup} shows the variation of the surface abundances of
$^{12}$C, $^{22}$Ne and $^{23}$Na during the same phases shown in the left panel.
An important phenomenon taking place during these phases is the formation of a
convective region in the helium burning shell, which eventually merges with the
penetrating envelope \citep{iben97, siess07}. The signature of this convective episode 
can be seen in the behaviour of the surface carbon: while in the phases following the 
beginning of SDU the surface $^{12}$C decreases, towards the end of SDU the amount of
carbon in the envelope increases, because the external convective region reaches layers 
touched by helium burning.

A further consequence of the deep penetration of the convective envelope, very important for
the present investigation, is the
increase in the surface $^{22}$Ne, whose behaviour is qualitatively similar to carbon.
During the first part of SDU the surface $^{22}$Ne decreases, as in the regions mixed with
the surface $^{22}$Ne was exposed to proton capture, which synthesized sodium. In the last
part of SDU the surface $^{22}$Ne increases, because the surface convection reaches 
regions of the star where repeated $\alpha$ captures favoured a significant synthesis 
of $^{22}$Ne, now transported to the surface. During the SDU the sum $^{22}$Ne$+^{22}$Na
increases by $70\%$; this will be extremely important for the sodium content of the
ejecta, considering that almost the totality of the $^{22}$Ne in the envelope is
converted into sodium during the TP phase.

The other two species touched by SDU (not shown in Fig.~\ref{f2dup} for clarity reasons) 
are oxygen, whose surface content decreases by $\sim 25\%$ (the oxygen abundance changes
from $0.0016$ to $0.0009$) and nitrogen, which increases by a factor $\sim 7$
(from $X(N)=4.98\times 10^{-5}$ to $X(N)=3.5\times 10^{-4}$).

\begin{figure*}
\resizebox{.33\hsize}{!}{\includegraphics{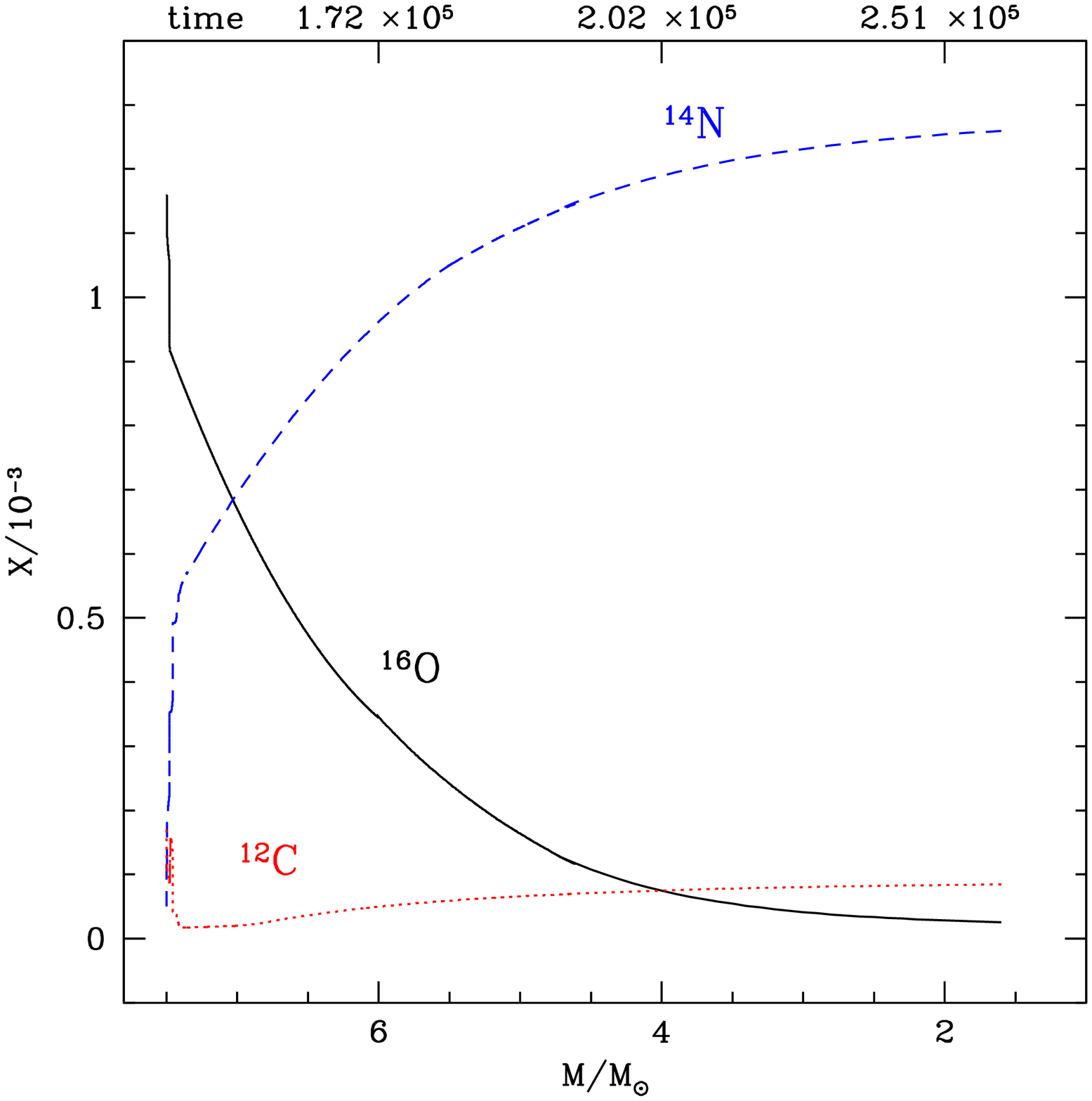}}
\resizebox{.33\hsize}{!}{\includegraphics{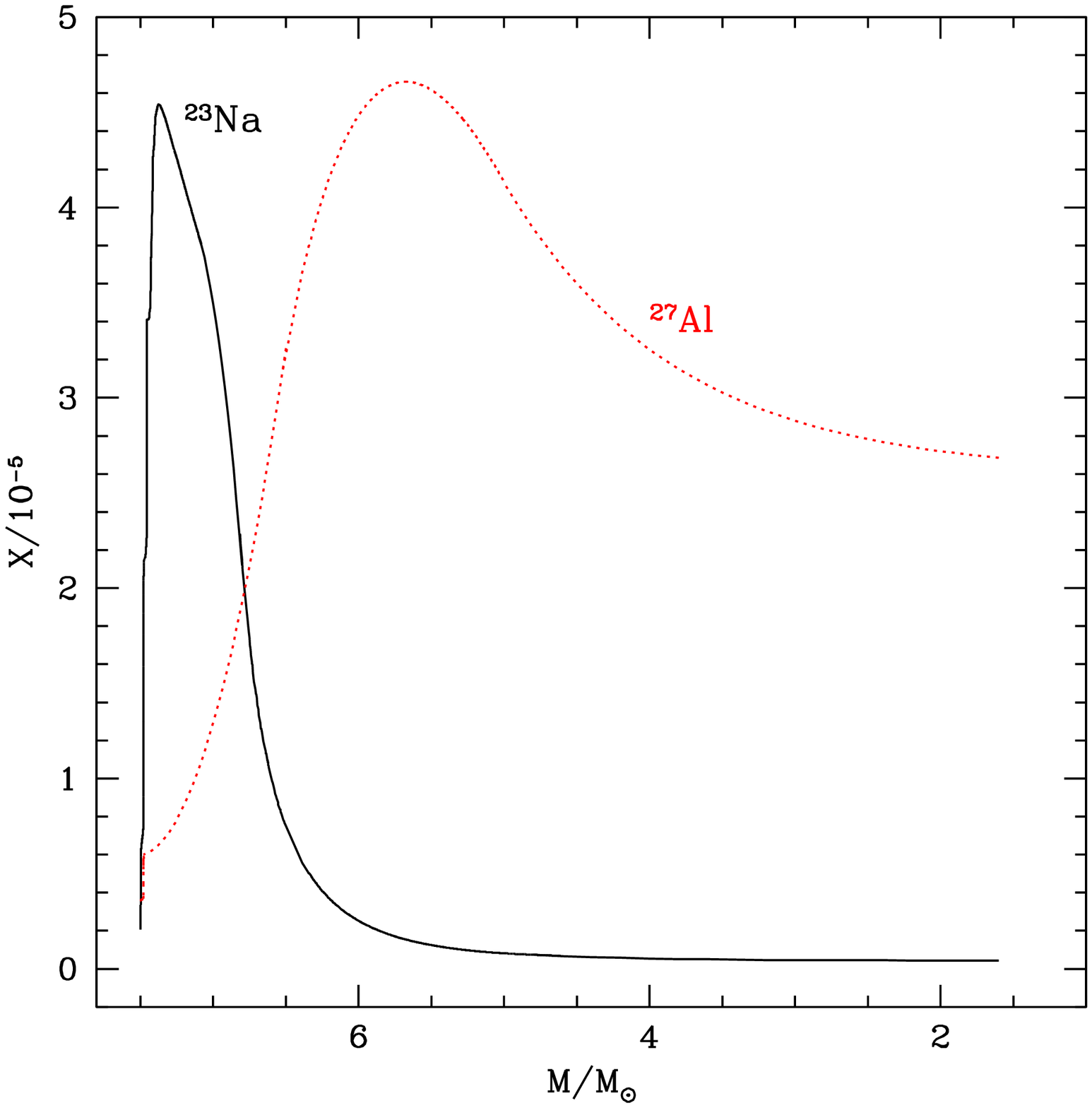}}
\resizebox{.33\hsize}{!}{\includegraphics{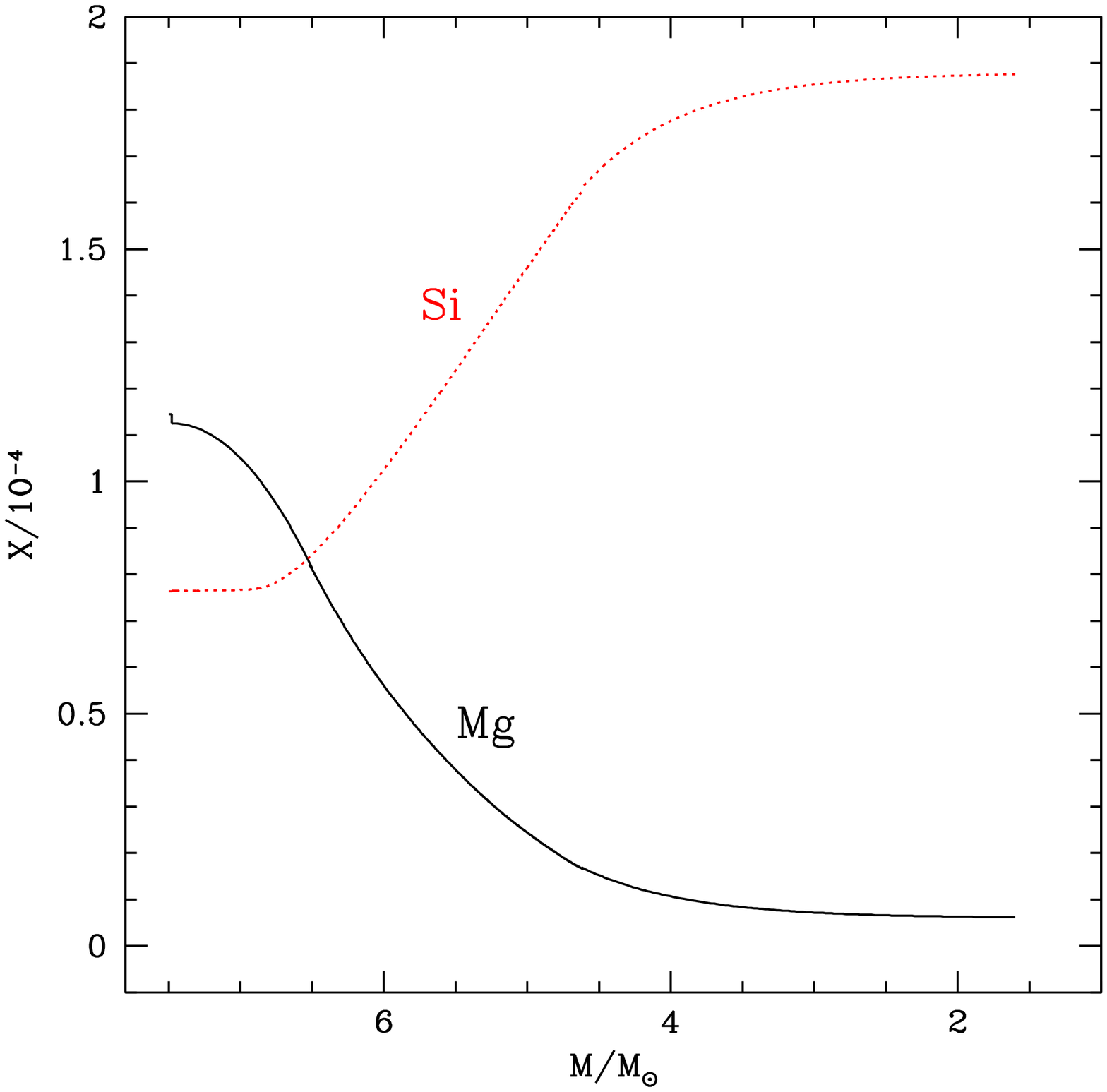}}
\caption{The evolution of the surface chemical composition of the $7.5~M_{\odot}$ model
shown in Fig.~\ref{f2dup}, during the TP phase. The mass fractions of the CNO elements
is shown in the top, left panel, whereas the abundances of sodium and aluminium are
reported in the top, right panel and those of magnesium and silicon in the
bottom, left panel.} 
\label{f75}
\end{figure*}

\subsection{The thermally pulsing phase}
\label{tp}
The envelope of SAGB stars is lost almost entirely during the
TP phase. Although extremely short, this is the evolutionary phase most important for the
feedback from these objects to the the interstellar medium, as it is now that most of the
gas and dust pollution occurs.

We show in Fig.~\ref{f75} the variation of the surface chemical
composition, in terms of the surface abundances of the CNO elements, sodium, and the
species involved in the Mg-Al-Si nucleosynthesis. The figure refers to the same model, of
initial mass $7.5~M_{\odot}$, shown in Fig.~\ref{f2dup}. We report 
the current mass of the star on the abscissa, to have a better idea of the chemistry of 
the gas ejected.

Massive AGB stars are characterized by strong HBB, since the early TP stages \citep{ventura11}. 
This holds
even more in metal-poor environments \citep{ventura13}. In the model reported in 
Fig.~\ref{f75} the temperature at the base of the envelope is slightly higher than 110 MK 
at the beginning of this phase and increases gradually during the following evolution, 
up to $\sim 140$ MK. Under these conditions the internal regions of the envelope 
experiment a very advanced p-capture nucleosynthesis.

As shown in the left panel of Fig.~\ref{f75}, the CNO species subject to the most 
significant
alterations are nitrogen, which increases by a factor of $\sim 4$ with respect to the
post-SDU abundance, and oxygen, which is almost entirely consumed. This behaviour is
a clear signature of the activation of the full CNO cycling at the base of the envelope.
The carbon trend is slightly increasing during this evolutionary phase, because the
carbon equilibrium abundance is higher the larger is the temperature at which CNO 
nucleosynthesis takes place.

The central  panel of Fig.~\ref{f75} shows that the Ne-Na nucleosynthesis is efficiently
activated since the beginning of the TP phase. The behaviour of sodium under HBB conditions 
was addressed in details by \citet{mowlavi99} and further tested in FST-based AGB models
by \citet{ventura06}. The present findings confirm on the qualitative side the results from
these earlier investigations, with the difference that in this case the accumulation of $^{22}$Ne
in the envelope, favored by efficient SDU, leads to a generally larger synthesis of sodium.
The sodium mass fraction during the initial TP-AGB phases reaches values
$\sim 10$ times higher than in the matter from which the star formed. After reaching a 
maximum, the abundance of sodium begins a negative trend, because the destruction channels,
particularly the $^{23}$Na$(p,\gamma)^{24}$Mg reactions, becomes more dominant
with respect to sodium production the higher is the temperature. 

The ignition of the Mg-Al-Si nucleosynthesis is witnessed by the behaviour of magnesium,
silicon (shown in the right panel of Fig.~\ref{f75}) and of aluminium
(middle panel). The development of this nuclear channel is rather complex, because 
the total magnesium is split into three isotopes, each exposed to proton capture.
The interplay among the various reactions was investigated in detail by \citet{arnould99},
while an exhaustive application to SAGB modelling is given in
\citet{siess08}. In a recent work \citet{ventura18} investigated the role of the
various cross-sections on the results obtained.
The destruction of $^{24}$Mg, the most abundant isotope, begins shortly after the activation 
of the Ne-Na nucleosynthesis, and continues efficiently for the whole TP phase, until 
$^{24}$Mg is almost exhausted. $^{25}$Mg and $^{26}$Mg are produced in the initial phases, 
then they are destroyed
by proton capture. The total Mg decreases steadily during the TP phase, until reaching final
abundances below $10\%$ of the original value. 

The destruction of magnesium favors the
production of aluminium and silicon. The behaviour of Al is not monotonic with 
time. Initially the surface aluminium increases up to values ten times higher than 
the initial content; in the final part the surface Al decreases, owing to the higher and 
higher efficiency of the $^{27}$Al$(p,\gamma)^{28}$Si reaction, which favours a significant 
increase in the surface silicon.

\subsection{Mass loss: the big unknown of the SAGB phase}
While it is generally recognized that strong HBB takes place in SAGB models
\citep{siess10, doherty14, doherty15},
the description of mass loss is still under debate and largely unknown.
The main issue in this context is that the mechanism favoring mass loss during the
TP-AGB phase is still poorly known.

The impact on the treatment of mass loss on the physical evolution of SAGB stars, on
the determination of the initial-final mass relationship and on the chemical yields
are discussed in \citet{doherty14, doherty15}.

In the previous works on this argument we calculated $\dot M$ via the \citet{blocker95} 
description \citep{ventura11, ventura13}, using the calibrated parameter
$\eta_R = 0.02$ for the whole TP phase. A consequence of this choice 
was that the gas ejected by SAGB models of mass $7-8~M_{\odot}$ is on the average less 
processed than in the 
lower mass counterparts, because the mass loss rates were so large that the rate at which the 
envelope is lost is faster than the pace of the p-capture reactions. This general 
behaviour holds for all the chemical species but helium, whose surface content, as discussed
earlier, is mainly determined by SDU.

\begin{figure}
\resizebox{1\hsize}{!}{\includegraphics{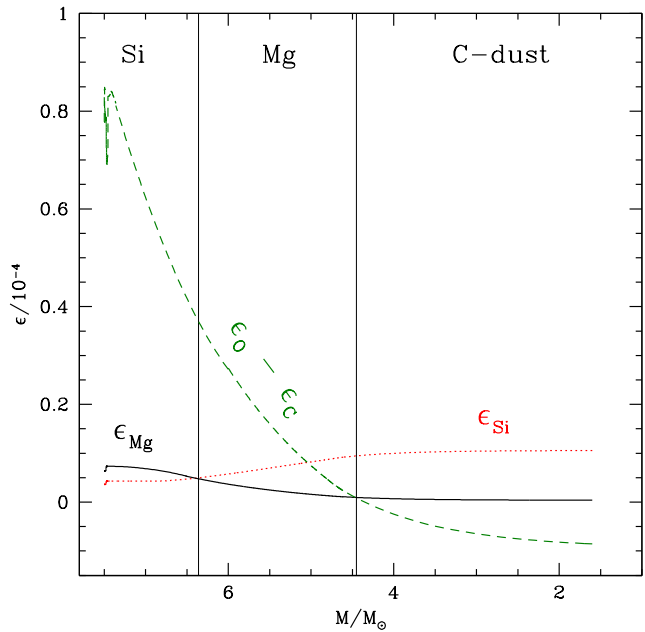}}
\caption{The number densities (relative to the hydrogen
particles) of magnesium (black solid), silicon (red dotted) and of the oxygen excess (green dashed) with respect to carbon at the surface of a M=7.5 $M_{\odot}$.
The three temporal phases are highlighted in each of which the behavior of mass loss is governed  by the abundance of Si, Mg C, respectively (see text).
}
\label{fdust}
\end{figure}

As discussed in section \ref{input}, we now reconsidered this choice, to account for the
effects of the radiation pressure on dust particles. To understand how this affects the
SAGB modelling, we show in Fig.~\ref{fdust} the evolution of
the surface number densities of the potential key elements for dust formation, namely
silicon, magnesium, and of the difference between the oxygen and carbon; for each
element $X$ we plot the quantity $\epsilon (X) = n(X)/n(H)$, i.e. the surface number
density, relatively to hydrogen \citep{fg06}.

During the initial TP-AGB phases the key element, determining the dust formation rate,
is silicon, which is the least abundant among the
species relevant to the formation of silicates; this situation is maintained until
the mass of the star decreases down to $\sim 6.5~M_{\odot}$. During this phase we 
expect that the mass loss rate increases, owing to the simultaneous increase in the
luminosity of the star and in the silicon content of the envelope; this is in 
agreement with the results obtained using the \citet{blocker95} recipe, which is
characterized by a steep dependence on the luminosity.\\
In the following phases magnesium replaces silicon as the key element to dust formation
(see Fig.~\ref{f75});
this is due to the destruction of magnesium and the synthesis of silicon by HBB,
which eventually makes the number of Mg particles smaller than Si. Because the surface 
density of the key species is now diminishing, less and less dust will be formed, with
the consequent decrease in the rate of mass loss (requiring use of
a smaller $\eta_R$).
We used this description until the $C/O$ ratio exceeds unity entering in the C-star phase, when the mass of the star drops below $\sim 4.5~M_{\odot}$ in  \citet{blocker95}; after which we kept the mass loss rate constant until the consumption of the whole envelope. The latter choice is somewhat arbitrary, but it has practically no effects on the gas yields.\\
It is also possible that mass loss driven by radiation pressure is resumed after a while, now stimulated by the
carbon dust grains formed.


\section{Discussion}
Table\ref{tabyield} reports the yields of SAGB stars of initial mass $6.5, 7, 7.5~M_{\odot}$.
The table includes the average chemical composition of FG stars and of SG stars 
belonging respectively to group P1 and E of NGC 2808 (see also Fig. 1).

\begin{table*}
\caption{ Average abundances of proton-capture elements in the extreme population E of 
NGC 2808 taken from Table 6 of Carretta et al. (2018) compared with the yields derived for 
SAGB of different masses.}                                       
\begin{tabular}{c c c c c c c c c}        
\hline
   &  Y  & $[C/Fe]$  &  $[N/Fe]$ &  $[O/Fe]$  & $[Na/Fe]$ &
$[Mg/Fe]$  &  $[Al/Fe]$ & $[Si/Fe]$   \\
\hline
& & & & NGC 2808 & & &  \\
& & & & stars & & &  \\
\hline
FG (P1) &  -  &  -  &  -  &  0.308 $\pm$ 0.058  &  -0.005 $\pm$ 0.067     &  0.384 $\pm$ 0.041  &  -0.114 $\pm$ 0.141 &  0.265 $\pm$ 0.026  \\
SG (E ) &  -  &  -  &  -  & -0.656 $\pm$ 0.161 &  0.592 $\pm$ 0.112  &  -0.050 $\pm$ 0.050    &  1.292 $\pm$ 0.113  &  0.390 $\pm$ 0.036  \\
\hline
& & & & SAGB yields & & &  \\
\hline
$M/M_{\odot}$  &  Y  & $[C/Fe]$  &  $[N/Fe]$  &  $[O/Fe]$  & $[Na/Fe]$ &
$[Mg/Fe]$  &  $[Al/Fe]$ & $[Si/Fe]$   \\
\hline
6.5 & 0.36 & -0.62  & 1.33 & -0.38 & 0.37 & -0.09 & 1.18 & 0.44  \\
7   & 0.36 & -0.60  & 1.33 & -0.33 & 0.44 & -0.04 & 1.07 & 0.48  \\
7.5 & 0.37 & -0.51  & 1.29 & -0.21 & 0.48 &  0.02 & 1.02 & 0.42  \\
\hline       
\hline     
\label{tabyield}
\end{tabular}
\end{table*}

\subsection{Direct influences of HBB: Magnesium depletion and Silicon enrichment}
Magnesium and silicon provide a direct information regarding the capability of the present
AGB models to reproduce the chemistry of SG stars, because their behaviour 
during the TP phase is monotonic (see Fig.~\ref{f75}). 

Mg is subject to destruction 
via p-capture: in the early TP phases $^{24}$Mg is destroyed in favour of $^{25}$Mg
and $^{26}$Mg, whereas in the more advanced TP evolution the heavier Mg isotopes are 
also exposed to efficient p-capture, which determines a fast decrease in the overall
surface magnesium. In the present SAGB models we find that the magnesium in the ejecta
is $[Mg/Fe]=0$, $2.5$ times smaller than FG stars. 

Silicon is synthesized by p-capture on Al nuclei. The production of silicon requires
not only that Mg burning is activated, but also that the process is sufficiently fast
to allow the accomplishment of the whole Mg-Al-Si chain on time scales shorter than 
(or comparable to) the evolutionary time scale; this requires large HBB temperatures,
in excess of $100$ MK. In the present computations, as reported in Table.\ref{tabyield},
we find that the silicon in the ejecta is $[Si/Fe]=+0.4$, to be compared with the
quantity measured in FG stars, i.e. $[Si/Fe]=+0.25$

\subsection{HBB equilibria: the behaviour of Aluminium and Sodium}
We find a significant Al enrichment in the ejecta, by a factor $\sim 10$ 
($[Al/Fe] \sim 1$). As discussed in section \ref{tp}, the Al 
content of the envelope during the TP evolution is determined by the
equilibrium between the production and the destruction channels, thus the results 
are extremely sensitive to the cross-section of the $^{27}$Al$(p,\gamma)^{28}$Si reaction: 
the comparison with the Al measured in E stars seems to require slightly lower rates, but 
the differences are not sufficiently significant to require a supplementary analysis on 
this argument. 

Sodium is the most delicate species entering this discussion, because the surface 
abundance reflects the equilibria between the production and the
destruction mechanisms (see Section \ref{tp}), both very sensitive to the temperature, 
thus exposed to
significant variation during the TP phase. This explains the TP evolution of the
surface sodium, shown in the top, right panel of Fig.~\ref{f75}. As reported in  
Table\ref{tabyield}, the average sodium of 
the ejecta, $[Na/Fe] \sim 0.5$, is similar to the quantity observed in E stars. Despite 
the vigorous destruction of sodium in the advanced TP phases, the ejecta are enriched in 
sodium (by a factor $\sim 3$), because during the initial part of the evolution
the sodium in the envelope reaches abundances $10-20$ times higher than
in the matter from which the star formed; this can be clearly seen in the top, right
panel of Fig.~\ref{f75}. Such a large increase in the surface sodium 
is partly due to the accumulation of $^{22}$Ne during the SDU (see discussion in 
section \ref{pretp}); an additional motivation is that the sodium
enrichment of the envelope takes place during the initial TP phases, when dust production 
is at the maximum efficiency (region labelled as Mg in the bottom, right panel of
Fig.~\ref{f75}): this favours the release of large quantities of sodium-rich gas 
into the interstellar medium.

\subsection{Can we obtain oxygen-free ejecta?}
Oxygen deserves a separate discussion. The SAGB yields reported in Table\ref{tabyield}
have $[O/Fe] \sim -0.2$, $\sim 4$ times smaller than FG stars. 
On the other hand, Fig.~\ref{fanti} above and Fig.~3 in \citet{carretta18} show that 
some stars belonging to group E have extremely low surface oxygen, up to 20 times lower 
than observed in FG stars; furthermore, some of these measurements are upper limits. 

We did several tests, by varying the HBB modelling and the description of mass loss:
our conclusion is the impossibility of producing SAGB ejecta 
with such low oxygen contents.
The only possibility is that the stars suffer extremely low mass loss during the
initial TP phases, until most of the oxygen in the envelope is consumed, via p-capture.
The rate of mass loss required to accomplish this task should be below
$10^{-6} \dot M/$yr, far too small for this high luminosity, expanded stars.
In this case the overall agreement between the yields of SAGB stars and the
chemistry of E stars would be substantially worse, because the gas ejected would be practically
sodium free and would show magnesium depletion and silicon enrichment far in excess
of the quantities observed.

On the other hand it is not possible to reproduce all the oxygen abundances of E stars
at the same time, given the large spread observed, amounting to almost 1 dex. Oxygen is 
very peculiar on this side, because Carretta et al. (20018 ) show that the range of measured values 
is much wider than the intrinsic error, $\sim 0.1$ dex. However the yields of the SAGB stars 
used here are compatible with the largest oxygen abundances among E stars. 

\subsection{Mixing on the RGB}
This difficulty of HBB  to account for the most extreme O--depletion cases was already clear a decade ago, as a very extended O--Na anticorrelation was already shown by NGC\,2808 giants \citep{carretta06} and by M\,13 giants \citep{sneden2004}. In this context, \cite{dantona07} proposed to explain it by extramixing during the giant stage, due to rotational evolution of  the stars born with the extreme anomalous composition provided by the AGB ejecta. \\
The stars we are analyzing are also extremely helium-rich (see Table 1), and during the evolution they do not develop the strong molecular weight barrier which prevents rotational mixing. For this reason, the first generation giants, born with primordial initial helium content, even if born with a similar rotational spread, would not be subject to the same kind of rotational mixing, and would not display a similar oxygen spread. Deep mixing, during which the bottom of the envelope
penetrates down until reaching the more internal regions of the H-burning shell, where CNO cycling had been active, was first invoked by \citet{denisenkov90} to explain results from high resolution spectroscopy of RGB stars in $\omega$ Centauri. 
We suggest now that this is a plausible explanation  for the minimum oxygen abundances, and possibly also for
the star-to-star differences  in the E population of NGC\,2808 as given in \citet{carretta18}. 
We studied the evolution of stars now evolving in the cluster, and formed with the composition of the SAGB ejecta of the $7.5~M_{\odot}$, i.e. $[O/Fe] = -0.2$\ and Y=0.37 (see Table~\ref{tabyield}), assuming that, during the
RGB phase, are exposed to non canonical extra-mixing, which favours a further depletion in the surface oxygen, but leaves the other elements signature of second generation, such as sodium and magnesium, unaltered. 
Because the gas is enriched in helium, these stars evolve faster than FG stars: consequently, for a given age, the mass currently evolving  through the RGB is lower. Assuming a 12 Gyr age for NGC 2808, a star formed directly from the SAGB winds, and nowadays experiencing  the RGB evolution, has a mass of $0.65~M_{\odot}$. 

\begin{figure*}
\resizebox{0.4\hsize}{!}{\includegraphics{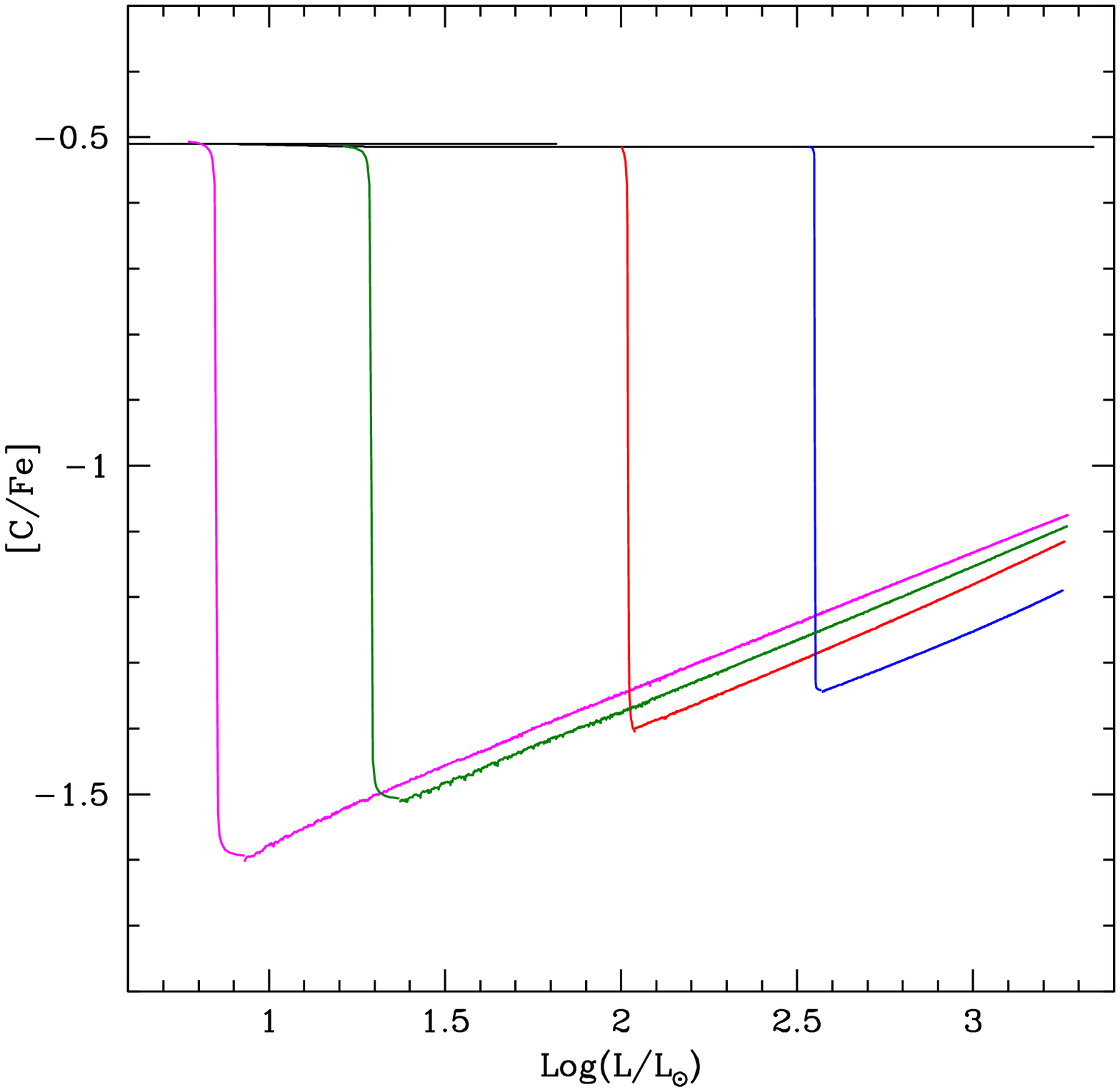}}
\resizebox{0.4\hsize}{!}{\includegraphics{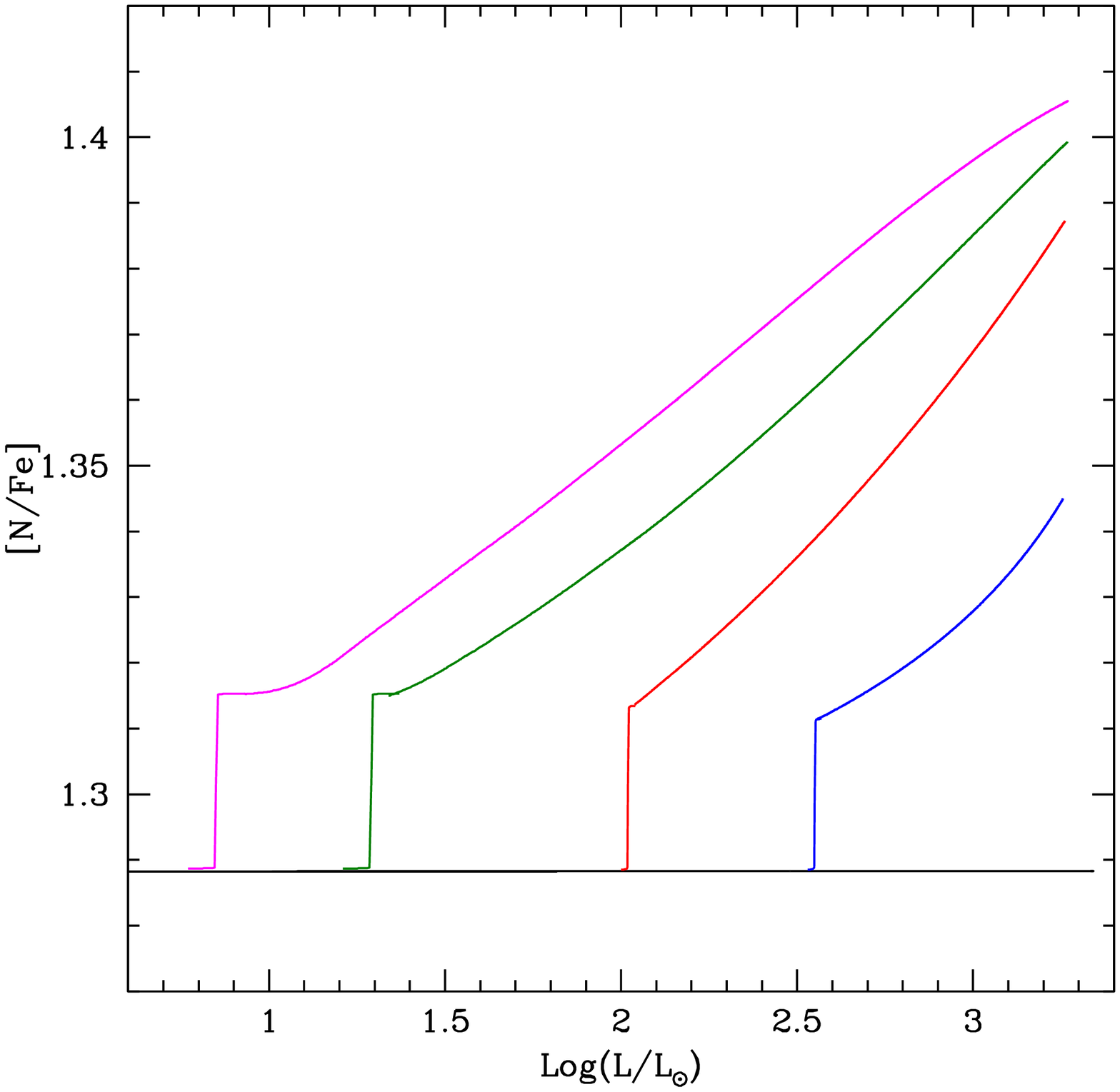}}
\resizebox{0.4\hsize}{!}{\includegraphics{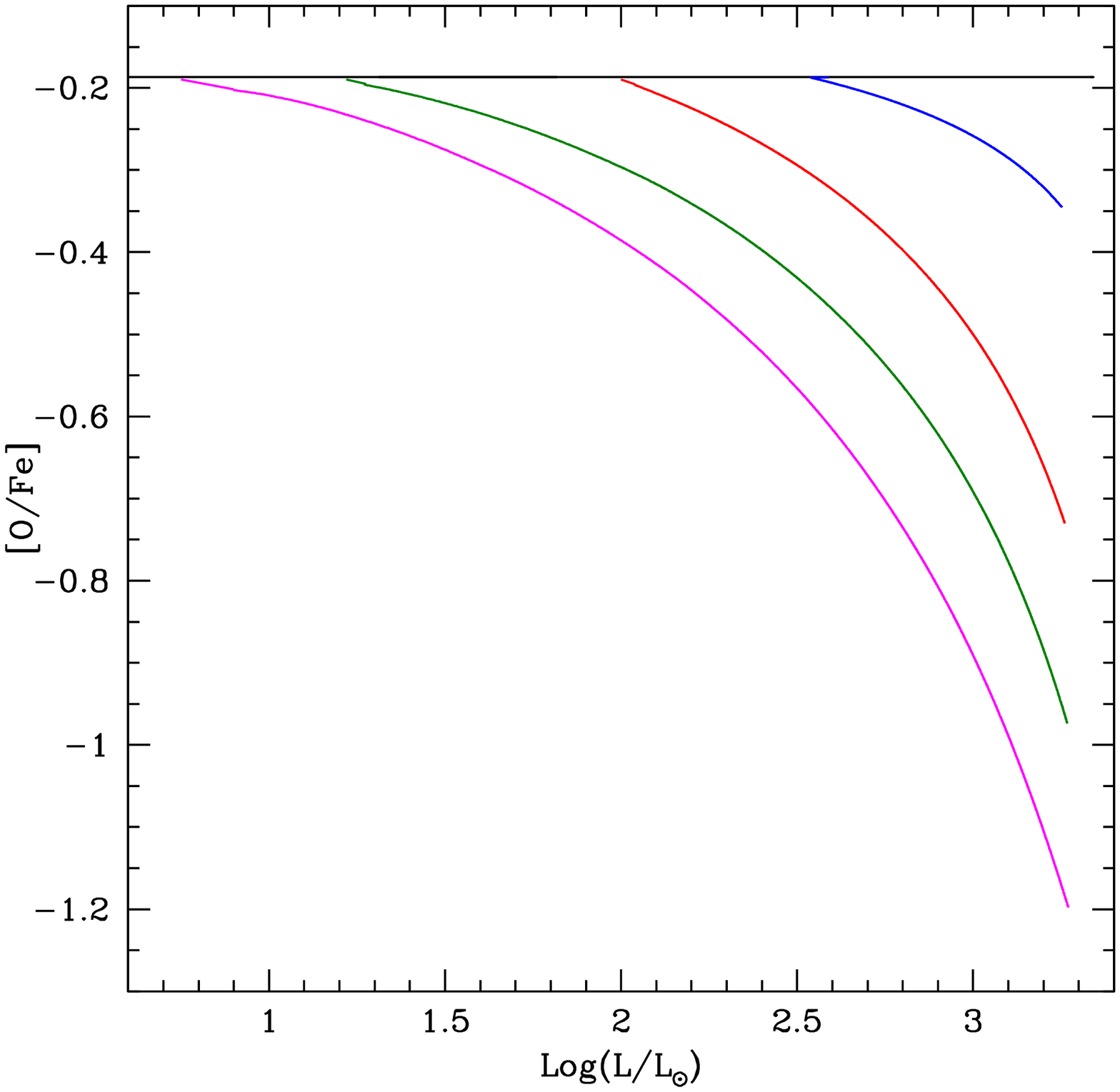}}
\resizebox{0.4\hsize}{!}{\includegraphics{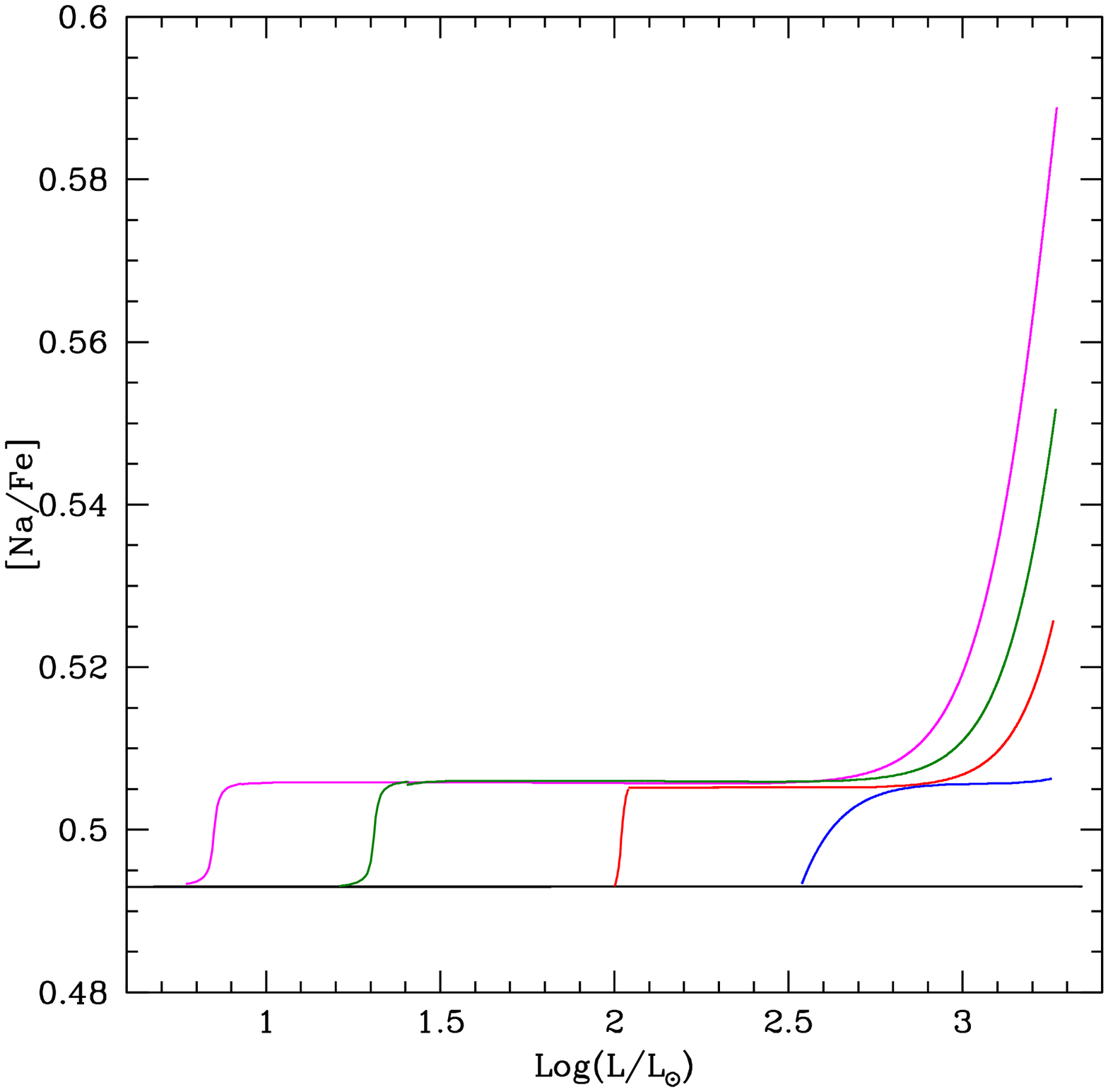}}
\caption{The variation of the surface mass fractions of the CNO elements and
of sodium during the RGB evolution of models of mass $0.65~M_{\odot}$ calculated
with the same chemical composition of the SAGB ejecta. The horizontal, black tracks
refer to the model calculated with canonical convective mixing, whereas the sequence
of magenta, green, red and blue lines refer to models where deep mixing was assumed 
starting from different evolutionary stages during the RGB ascending.
}
\label{frgb}
\end{figure*}


We calculated various evolutionary sequences of $0.65~M_{\odot}$ models, with the 
chemical composition of the SAGB ejecta reported in Table~\ref{tabyield}. 
We assumed non canonical, deep mixing, started at different stages of the RBG phase, and prosecuted until the helium flash. To model such mixing we assumed that the surface convective zone penetrates inwards, down to regions whose temperature is $\Delta \log T = 0.05$ hotter than the formal border, fixed by the Schwarzschild criterion. In the extra-mixed zone we assumed the exponential decay of the convective velocities, as described in section 4.1.
The results are shown in Fig.~\ref{frgb}, where the variation of the surface abundances of the
individual species are shown as a function of the luminosity of the star. We do not
show the behaviour of Mg, Al and Si, because no variation in the surface mass 
fractions of these elements is found.

The top, left panel of Fig.~\ref{frgb} shows that deep mixing causes a decrease 
in the surface carbon, whose initial abundance is $\sim 3$ times smaller than FG stars. 
Note that after the initial, dramatic drop in the carbon content, the $^{12}$C mass
fractions starts to increase, because the base of the envelope reaches layers touched
by full CNO cycling, where the equilibrium $^{12}$C is higher than in CN burning
regions.

The surface nitrogen, shown in the top, right panel of Fig.~\ref{frgb}, is subject to
a modest increase, below $0.1$ dex. The reason for this behaviour is that the initial
N is almost $20$ times higher than in FG stars, which leaves little room for a further,
percentage increase in the N content (note that the quantity reported on the y-axis is
in logarithmic scale). 

We see in the bottom, left panel of Fig.~\ref{frgb} that oxygen is 
exposed to significant alteration during the RGB ascending. The gradual penetration of
the surface convection, down to regions of the star touched by CNO nucleosynthesis, 
determines a progressive reduction of the surface oxygen, which can be depleted
by a factor $\sim 10$ during the RGB phase (note that the initial oxygen was
4 times smaller than FG stars). 

The behaviour of sodium, shown in the bottom, right panel of Fig.~\ref{frgb}, is 
qualitatively similar to nitrogen. Despite the surface convection reaches zones involved in
Ne-Na nucleosynthesis, the $^{22}$Ne available is extremely small and
the initial sodium is 3 times higher than in FG stars; these conditions prevent
a significant, percentage increase in the sodium content, thus [Na/Fe] raises by
less than 0.1 dex.

\subsection{Understanding the most contaminated stars in NGC 2808}
The discussion in section \ref{2808} outlined the importance of NGC 2808 for the
comprehension of the formation of multiple populations in GC. We stressed how important
is the interpretation of the stars with the most extreme chemical composition
(E stars, in Carretta et al. 2018), to understand whether self-enrichment by AGB stars
might explain the chemical abundances observed: it is essential that the chemical
composition of these stars are in agreement with the ejecta from SAGB stars.
Fig.~\ref{fanti} shows the comparison between the yields of the SAGB models discussed
here and the chemical composition of E stars in the O-Na and Mg-Al planes.

The magnesium depletion ($\delta Mg = -0.4$) and the silicon enrichment 
($\delta Si = +0.2$) reported in Table ~\ref{tabyield} are in nice agreement with the 
observations: this is a confirmation that the HBB temperatures are sufficiently hot to 
allow the nucleosynthesis required to favours the destruction of the magnesium isotopes 
and the synthesis of significant quantities of silicon\footnote{We remark here that 
silicon is among the most abundant species, among the various elements: an increase in the
surface silicon of $+0.2$ dex indicates that significant amounts of magnesium (and 
aluminium) are converted into silicon}.

The aluminium and sodium abundances of E stars are also in good agreement with the chemistry 
of the SAGB ejecta. This finding indicates that the interplay of HBB and mass loss 
during the SAGB phase is appropriate to allow the required enrichment in Na and Al.
Sodium is particularly indicative on this side, because its behaviour is subject to significant
changes during the TP phase: producing gas enriched in sodium requires a strong mass
loss during the initial TP phases, when the surface regions of the stars are
characterized by large sodium abundances. 

The nucleosynthesis at which the base of the envelope is exposed favours the release
of oxygen-poor gas. As shown in Fig.~\ref{fanti}, we find that the average oxygen in the 
SAGB ejecta is $[O/Fe]=-0.2$, consistent with the largest abundances measured in stars in 
group E. To explain the most extreme Oxygen abundances measured in He-Extreme stars we 
propose that deep mixing occurs during the RGB evolution. This would be, in principle, 
the consequence of the chemical mixing associated with angular momentum evolution of 
initially rotating stars. The large spread in these values may be imputed to different 
reasons. First of all,  the stars with the smallest oxygen content may be those into 
which deep mixing started earlier during the RGB evolution, as displayed in Figure 3; 
another possibility is that the stars were born with somewhat different AGB yields for 
oxygen (as displayed in Table 1) and they suffer a similar extramixing, but end up with 
a different surface oxygen.

\section{Conclusions}
We investigate the formation of multiple populations in the globular cluster NGC 2808. 
On the wake of previous explorations, we focus on the possibility that the
SGs  formed with the contribution of the gas ejected by the progeny of stars of mass 
$4~M_{\odot} < M < 8~M_{\odot}$, during the AGB phase.

This analysis is extremely timely, because recent data from high resolution spectroscopy
present a complete overview of the chemical composition of dozens of FG and SG stars in
the cluster, making available the individual abundances of all the light elements: 
oxygen, sodium, magnesium, aluminium an silicon.

These observational results offer an unprecedented opportunity of constraining the 
physical conditions and the degree of nucleosynthesis at which the contaminated gas, 
from which SG stars formed, was exposed. This is particularly important for the AGB 
self-enrichment scenario, which we discuss here, because according to this hypothesis the 
stars with the most contaminated chemistry should have the same chemical composition of 
the most massive stars escaping type II SNe explosion, i.e. $6.5-8~M_{\odot}$ stars, 
undergoing the SAGB evolution. 

We present new models for massive AGB stars, evolved from the same chemical
composition of the stars belonging to the FG of the cluster. We propose a modification
of the mass loss description used in previous works, to consider the decrease in the
effects of the radiation pressure, when magnesium is destroyed by HBB at the base of the 
envelope.

Our results show that the present models nicely reproduce the depletion of magnesium
and the enrichment in aluminium and silicon observed in the most contaminated stars.
Furthermore, unlike previous explorations, the predictions from this modelling can 
nicely reproduce the sodium enrichment observed, despite the intense sodium
destruction occurring in the advanced AGB phases. This agreement is due to the combined 
effects of: a) the accumulation in the envelope of $^{22}$Ne, synthesized in the regions
exposed to $3\alpha$ nucleosynthesis, which occurs during the second dredge-up; b) the
strong mass loss suffered by these stars during the phase with the maximum enrichment in
sodium.

The large spread in the oxygen abundances is explained by invoking deep mixing during
the RGB phase. The stars with the largest oxygen $[O/Fe]=-0.2$, in agreement with the
results from SAGB modelling, did not suffer any extra-mixing; conversely, the stars
with the lowest abundance of oxygen are those exposed to deep mixing since the early
RGB phases.

Concluding  we agree with Carretta et al.(2018) that NGC 2808 is one of the best 
benchmark to test any scenario for the origin and the evolution of multiple populations in 
GC  but we are not equally in agreement with the fact that large inconsistencies, related 
to the nucleosynthesis exist  when the temporal  sequence for the formation of the various 
populations is formulated within the self-enrichment scenario by massive AGB. 
In particular, in this work the extreme He-rich population allows us to test AGB scenario 
focusing on the role of very massive AGB stars. In this evolutionary phase the great unknown 
is  the  mass loss, but we have shown that a Blocker's approach with  a calibration that 
takes into account the variation of Mg during the evolution in AGB nicely reproduces the 
observe depletion of magnesium and the enrichment in aluminium observed in Extreme He 
enriched stars. In addiction taking into consideration the deep mixing in RGB we are also 
able to reproduce the large spread observed in the Oxygen abundance. This last hypothesis 
will  be easily tested when the measurements of the Oxygen abundance will be available also 
in main sequence stars of NGC2808 where we expect larger and less dispersed average values   
than those actually observed in RGB.

\section*{Acknowledgments}
We thank the referee for his/her useful comments and  Eugenio Carretta for sharing with us  Carretta et al. (2018)'s tables before on-line publication. Flavia Dell'Agli  acknowledges support provided by the Spanish Ministry of Economy and Competitiveness (MINECO) under grant AYA-2017-88254-P.
Marco Tailo acknowledges support by the European Research Council through the ERC-StG 2016, project 716082 `GALFOR'(http://progetti.dfa.unipd.it/GALFOR/) and by the MIUR through the the FARE project R164RM93XW `SEMPLICE'.

\end{document}